\documentclass[pr,twocolumn,showpacs,amsmath,nobibnotes,nofootinbib,floatfix,aps]{revtex4-1}
\usepackage{color}
\usepackage{ulem}
\usepackage{graphicx}
\usepackage{epstopdf}
\usepackage{bm}
\usepackage{amssymb}
\usepackage{amsmath}
\usepackage[unicode=true,colorlinks=true,citecolor=blue,urlcolor=blue]{hyperref}
\usepackage{epsfig}

\begin{document}

\title{Circular photon drag effect in bulk tellurium}
\author{V.\,A.\,Shalygin$^{1,2}$,
M.\,D.\,Moldavskaya$^{1}$, S.\,N.\,Danilov$^2$,  I.\,I.\,Farbshtein$^3$, and L.\,E.\,Golub$^3$
}
\address{$^1$Peter the Great St. Petersburg Polytechnic University, 
St. Petersburg 195251, Russia}
\address{$^2$Terahertz Center, University of Regensburg, 
Regensburg 93040, Germany}
\address{$^3$Ioffe Institute, 
St. Petersburg 194021, Russia }

\date{\today}

\begin{abstract}
The circular photon drag effect is observed in a bulk semiconductor. The photocurrent caused by a transfer of both translational and  angular momenta of light to charge carriers is detected in tellurium in the mid-infrared frequency range. Dependencies of the photocurrent on the light polarization and on the incidence angle agree with the symmetry analysis of the circular photon drag effect. Microscopic models of the effect are developed for both intra- and inter-subband optical absorption in the valence band of tellurium. 
The shift contribution to the circular photon drag current is calculated. An observed decrease of the circular photon drag current with increase of the photon energy is explained by the theory for  inter-subband optical transitions. Theoretical estimates of the circular photon drag current  agree with the experimental data.
\end{abstract}


\maketitle

\section{Introduction}

Circularly polarized light has an angular momentum which
can be transferred to charge carriers. The transfer of angular momentum from 
light to charge particles is studied in various fields of condensed matter physics, 
in particular in   semiconductors and metals~\cite{Dyakonov,Zutic}. 
Interaction of circularly polarized light with matter results in spin orientation 
of free carriers~\cite{Kusr_Landw}, in magnetization control by light in magnetic materials~\cite{Kimel_RMP_2010}, and in generation of charge and spin 
currents in semiconductor and ferromagnetic structures~\cite{GanichevPrettlreview,DyakonovIvchenko,spin_current_book}
and graphene~\cite{GlazovGanichevReview,graphene_Floquet,Katsnelson_2015}.

Presence or absence of the light angular momentum depends on its polarization state. 
However, light wave always carries a translational momentum. First demonstrated by 
Lebedev in his classical experiment~\cite{Lebedev}, light translational momentum has 
been detected in semiconductors by an electrical current. The current is generated 
by the transfer of the translational momentum from light to charge carriers, 
see  e.g. Refs.~\cite{EL_book,Ganichev_Prettl_book}
and references therein. The photocurrent generation due to this mechanism is known as the 
photon drag effect. It is investigated as a fundamental phenomenon in various media, 
e.g. in semiconductors~\cite{Ganichev_Prettl_book} and dielectrics~\cite{PDE_diel}, 
and has practical application in work of 
photodetectors~\cite{Rogalski_book,Ganichev_Prettl_book,Helicitydetector,Bruederman2014}. 
Now the photon drag effect is intensively studied in graphene for material 
characterization~\cite{GlazovGanichevReview} and generation of terahertz 
radiation~\cite{PDE_graphene_France,PDE_graphene_Obraztsov_Sci_Rep,PDE_graphene_Obraztsov_PRB}, 
in carbon nanotubes~\cite{PDE_SWNT}, and in thin metal films, where the photon drag 
current enhances in vicinity of surface plasmon resonance~\cite{surf_plasmon_drag}.

An additional control over the photon drag current can be made at simultaneous transfer 
of both angular and translational momenta from light to charge carriers. There is 
a part of the photon drag current which is sensitive to the light helicity and reverses 
its direction at switching from right- to left-circular polarization of light. 
Generation of the helicity-dependent photon drag current is known as a circular photon 
drag effect (CPDE). In contrast to the intensively studied photon drag 
currents insensitive to the circular polarization, the history of CPDE is rather short. While 
CPDE has been discovered theoretically  already in 1980's 
in Refs.~\cite{IP_1980,Belinicher_1981}, and some theoretical research has been made in 
the next decade~\cite{CPDE_cylinders}, 
experimentally it has been  demonstrated only in 2007 by studying 
quantum wells~\cite{CPDE_110_QWs,CPDE_110_QWs_2}.
Further experiments also deal exclusively with
two-dimensional 
 systems as photonic crystal slabs~\cite{Hatano}, 
graphene~\cite{graphene_ac_Hall,graphene_CO2,graphene_restrahl},  metamaterials~\cite{CPDE_metamat}, 
and quantum-well structures~\cite{CPDE_CR_Budkin}. Most recently CPDE become in focus of 
investigations of two-dimensional surface states in  topological insulators~\cite{McIver,TI_PRL}  having an aim to realize 
optical control of spin currents and characterize high frequency electron transport in these novel materials.
However, CPDE has not been observed so far in bulk systems.
The reason is that CPDE is forbidden by symmetry in cubic crystals like III-V semiconductors, or it is masked by other effects in these media where it is allowed.
In this work, we address the fundamental question: whether a transfer of 
both angular and translational momenta is possible in three-dimensional structures?

We report on the observation of CPDE in a bulk semiconductor demonstrating that the photon drag current sensitive to the light helicity is indeed possible in three-dimensional systems. For this purpose we choose tellurium 
which
demonstrates a few related phenomena, namely, 
electric current induced optical activity, linear photon drag effect, linear and circular 
photogalvanic effects~\cite{Ganichev_Prettl_book,Te_2012}. 
In contrast to the two-dimensional systems~\cite{CPDE_110_QWs,CPDE_110_QWs_2}, for tellurium  we can choose a particular geometry where CPDE is not hidden by any other effect. We show that the values of CPDE current is two orders of magnitude higher than in quantum well structures.

The paper is organized as follows. In Sec.~\ref{Sec:phenom}, the phenomenological analysis of photocurrents in  tellurium is preformed. In Sec.~\ref{Sec:observation} we present results of experimental observation of the CPDE current.
In Sec.~\ref{Sec:theory}, the microscopic theory of CPDE is developed. Section~\ref{Sec:discussion} discusses the obtained results, and Sec.~\ref{Sec:conclusion} concludes the paper.

\section{Determination of experimental geometry}
\label{Sec:phenom}

\begin{figure}[t]
\includegraphics[width=0.7\linewidth]{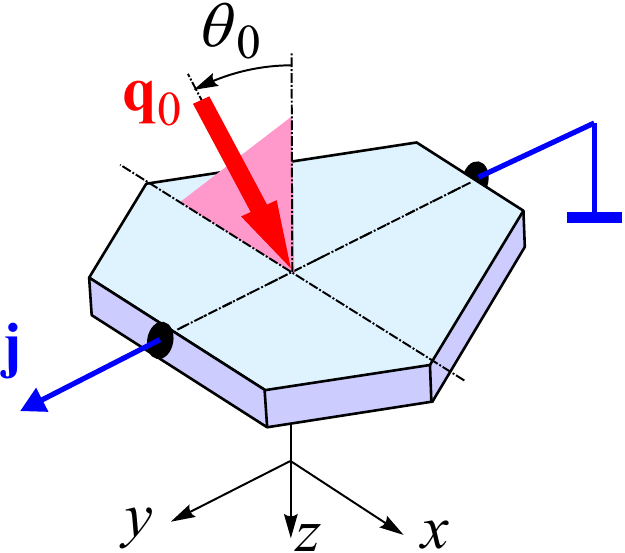}
\caption{The experimental geometry.}
\label{fig:exp_geom}
\end{figure}

In order to choose a proper geometry for observation of CPDE current we perform the symmetry analysis of helicity-dependent photocurrents in tellurium.
The point symmetry group of tellurium is D$_3$. In the plane perpendicular to the optical axis $z$, there are three rotation axes $C_2'$, which form an angle of $120^\circ$ with each other. We denote one of them as $x$, and a perpendicular axis in the same plane as $y$, see Fig.~\ref{fig:exp_geom}.
We consider a radiation incident in the plane $(xz)$. Performing a symmetry analysis we obtain the photocurrent which reverses its direction under switching from right-hand to left-hand polarized radiation. The density of this photocurrent, $\bm j^{circ}$, proportional to the circular polarization degree of light $P_{circ}$ is given by
\begin{align}
\label{j_z_phenom}
&	j_z^{circ} = \gamma P_{circ} {q_z \over q} E^2, \\
&	j_x^{circ} = \tilde{\gamma} P_{circ} {q_x \over q} E^2 + \tilde{T} P_{circ} {q_x^2 \over q} E^2, \\
&	j_y^{circ} = T P_{circ} {q_x q_z \over q} E^2.
\label{j_phenom}
\end{align}
Here $\bm q$ and  $\bm E$ are the radiation wavevector and electric field, respectively, and $E=|\bm E|$. 
The constants  $\gamma$ and $\tilde{\gamma}$ describe the circular photogalvanic effect caused solely by transfer of an angular momentum of photons to free carriers but not accompanied by a linear momentum transfer.  
The longitudinal CPDE current described by the constant $\tilde{T}$ 
is present due to a trigonal symmetry of tellurium. The transverse  CPDE described by the constant $T$ is caused by a non-equivalence of the $z$ direction and the  directions in the perpendicular plane $(xy)$, i.e. due to uniaxiality of tellurium. This 
CPDE current is odd in the incidence angle $\theta_0$.

The above Eqs.~\eqref{j_z_phenom}-\eqref{j_phenom} demonstrate that the current $j_y^{circ}$ is caused solely by the CPDE in contrast to two other photocurrent components. Therefore in the experimental part we focus on the photocurrent transverse to the incidence plane $(xz)$.

\section{Observation of CPDE}
\label{Sec:observation}

Experimental investigations were performed on a $p$-Te single crystal, which is characterized at room temperature by the concentration $p = 7 \times  10^{16}$~cm$^{-3}$ and the hole mobility $\mu= 700$~cm$^2$/(V s). The tellurium single crystal was grown by the Czochralski method in a hydrogen atmosphere. 
The samples had the form of a hexagonal prism, in which the lateral surface was a natural facet of the crystal,
and the end faces were subjected to optical polishing. It was found that the crystal under investigation exhibits a natural optical activity and is laevorotatory.
The thickness of the sample in the direction of the crystal optical axis $z$ was $L = 0.8$~mm.
To measure the  photocurrent $J_y$ in the direction  perpendicular to the incidence plane, two contacts were located at the lateral surface of the sample. The contacts were prepared from an alloy of tin, bismuth, and antimony with a low melting temperature (Sn : Bi : Sb = 50 : 47 : 3). 
A few samples fabricated from the same single crystal were studied demonstrating similar results.

We applied mid-infrared radiation of a tunable Q-switched as well as pulsed
TEA CO$_2$ lasers with an operating spectral range from 9.2 to 10.8~$\mu$m corresponding 
to photon energies ranging from 114 to 135~meV~\cite{laser1,laser2}. The laser pulses
had the peak power $P$ of about 500~W (for TEA laser after attenuation), the
pulse duration 100-250~ns, and the repetition frequency up to 160~Hz for
Q-switch and near 1~Hz for TEA laser. The radiation power was controlled by a
room-temperature photon drag~\cite{GanichevPDdetector,pddetector} and mercury cadmium telluride detectors. 
The radiation was focused in a spot of 0.5~mm diameter being much smaller than the
sample sizes in the $x$- and $y$ directions (8 and 7~mm, respectively).
The spatial beam distribution has an almost Gaussian profile, 
measured by a pyroelectric camera~\cite{Ganichev1999,Ziemann2000}.
The photocurrent  was measured by means of a storage oscilloscope.

The upper end face of the sample was illuminated by a laser beam under incidence angle $\theta_0$.
The latter is defined as the angle between the wavevector of the incident radiation $\bm q_0$  and the $z$-axis. The angle $\theta_0$ shown in Fig.~\ref{fig:exp_geom} is positive.
The incidence plane $(xz)$ contains the crystallographic axis  $x$ being of the two-fold rotation axis $C_2'$.
The laser radiation was linearly polarized. By applying a Fresnel $\lambda/4$ rhomb we modified the radiation polarization from linear to elliptical. The circular polarization of the light at the Fresnel rhomb output,  $P_{circ}^0$, was varied from -1 (left-handed circular polarization $\sigma_-$) to +1 (right-handed $\sigma_+$) according to $P_{circ}^0 = \sin {2\varphi}$, where $\varphi$ is the azimuth of the Fresnel rhomb. 

\begin{figure}[t]
\includegraphics[width=0.52\linewidth]{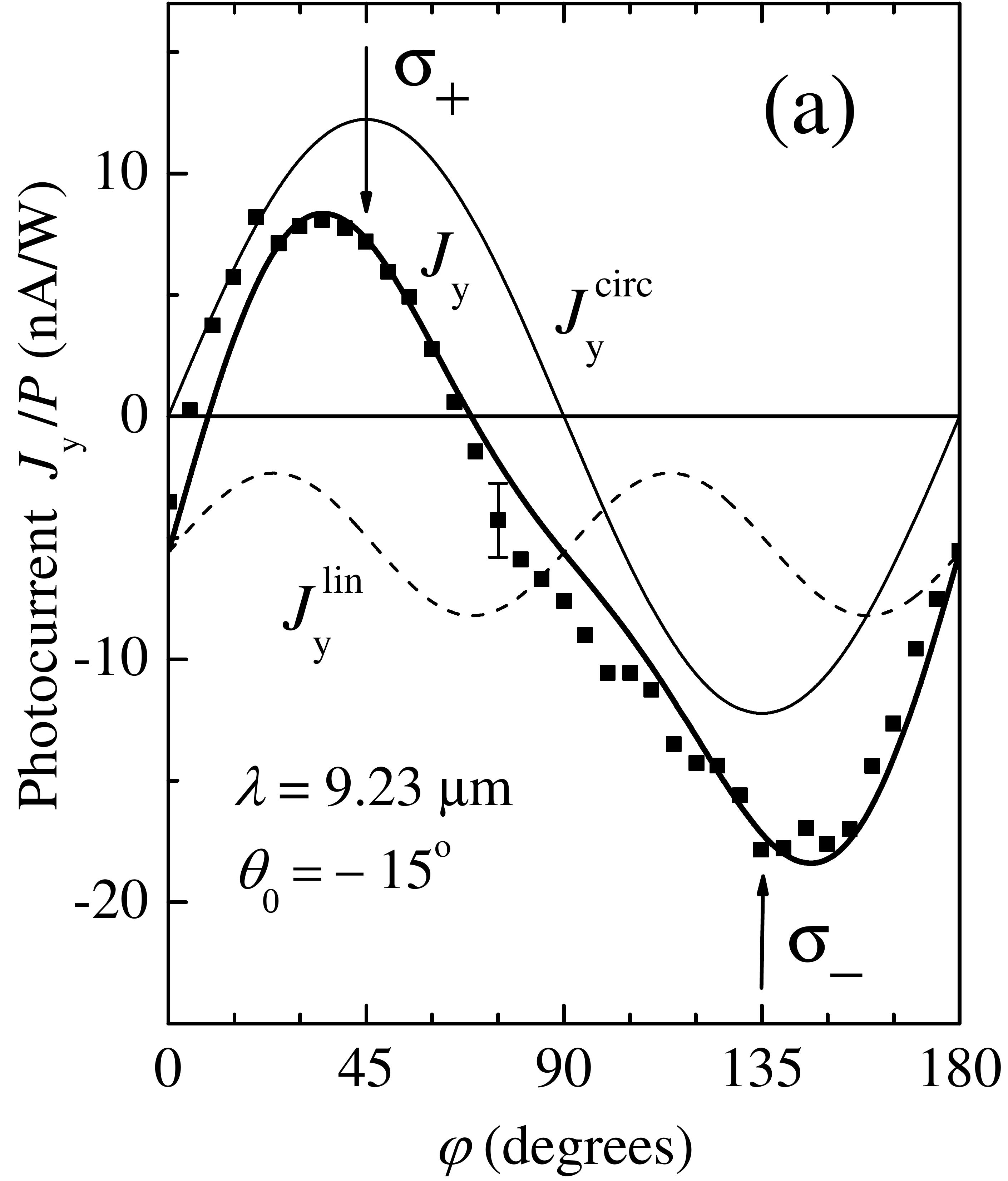} \:
\includegraphics[width=0.44\linewidth]{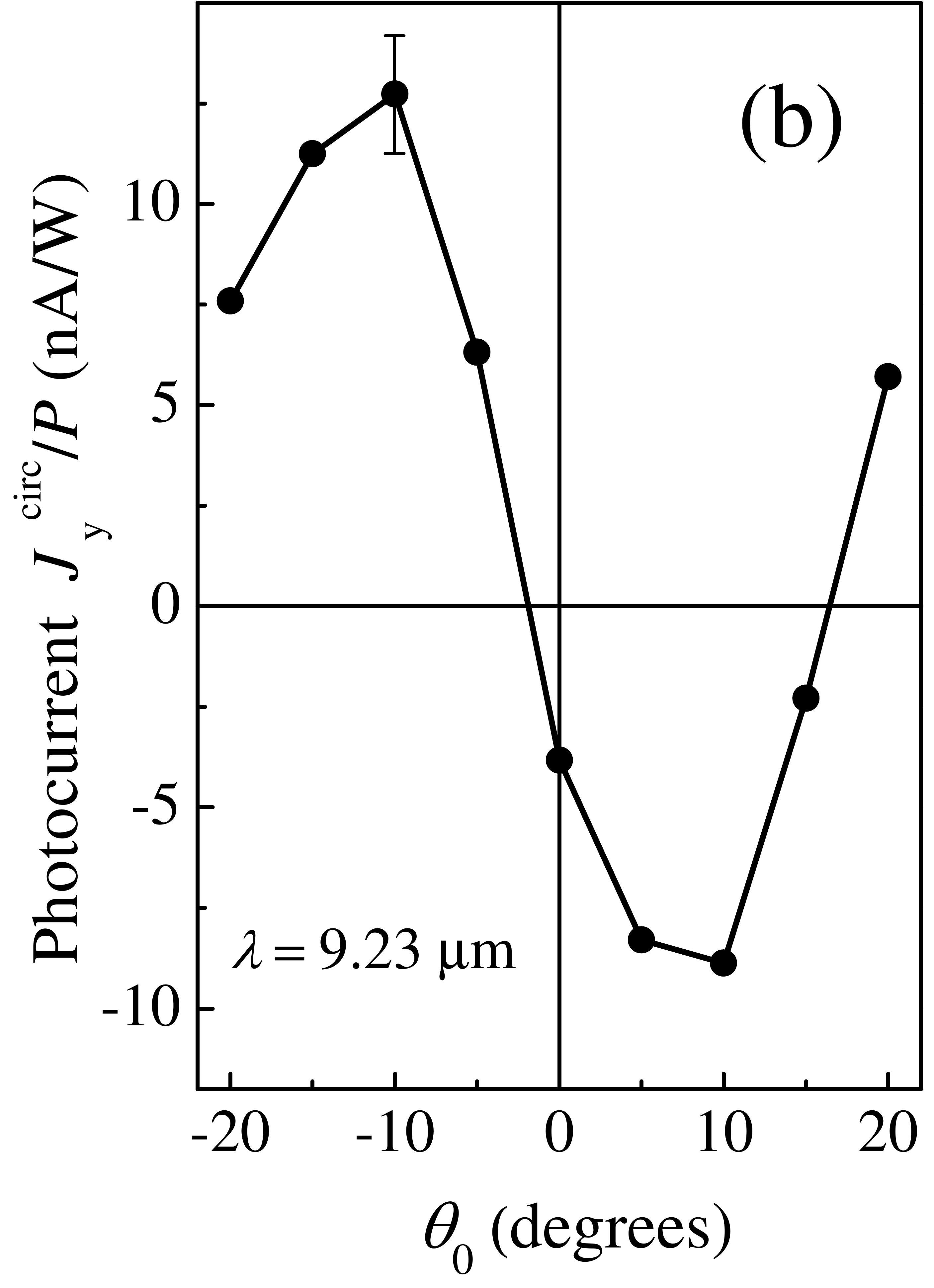}
\caption{(a) The transverse photocurrent dependence on the Fresnel rhomb rotation angle. $J_y^{circ} \propto  \sin{2\varphi}$ and $J_y^{lin}$ represent the circular and linear photocurrents, respectively.
(b) The dependence of the circular photocurrent on the incidence angle.}
\label{fig:phi-dependence}
\end{figure}

Figure~\ref{fig:phi-dependence}(a) shows a typical dependence of the transverse photocurrent $J_y$ normalized to the  laser power $P$ on the Fresnel rhomb azimuth under oblique incidence of the laser beam (at $\theta_0=-15^\circ$). 
The photocurrent at each azimuthal angle $\varphi$ depends linearly on the laser power $P$.
The experimental data are well described by the following phenomenological expression:
\begin{equation}
\label{E1}
J_y = C	\sin {2\varphi} + L_1 \sin {4\varphi} + L_2,
\end{equation}
where the first term is proportional to $P_{circ}^0$ and corresponds to the ``circular'' photocurrent $J_y^{circ}$ which we are interested in. Two other terms represent the ``linear'' photocurrent $J_y^{lin}$ which appears under elliptically polarized excitation.
The linear photocurrent is insensitive to the radiation helicity.
The photocurrent has a substantial part dependent on the light helicity. Figure~\ref{fig:phi-dependence}(a) shows that the photocurrent has opposite directions at excitation by right-handed ($\sigma_+$) and left-handed ($\sigma_-$) polarized light. 

In order to extract the current sensitive to the light helicity, ${J_y^{circ}=C	\sin {2\varphi} }$, we perform 
analysis of the $\varphi$-dependencies of the transverse photocurrent $J_y$ at various incidence angles. This allows us to reveal the dependence of the circular photocurrent amplitude $C$  on $\theta_0$. It is presented in Fig.~\ref{fig:phi-dependence}(b). One can see that the circular photocurrent is mainly an odd function of $\theta_0$ with an admixture of a small even contribution.

\begin{figure}[h]
\includegraphics[width=0.9\linewidth]{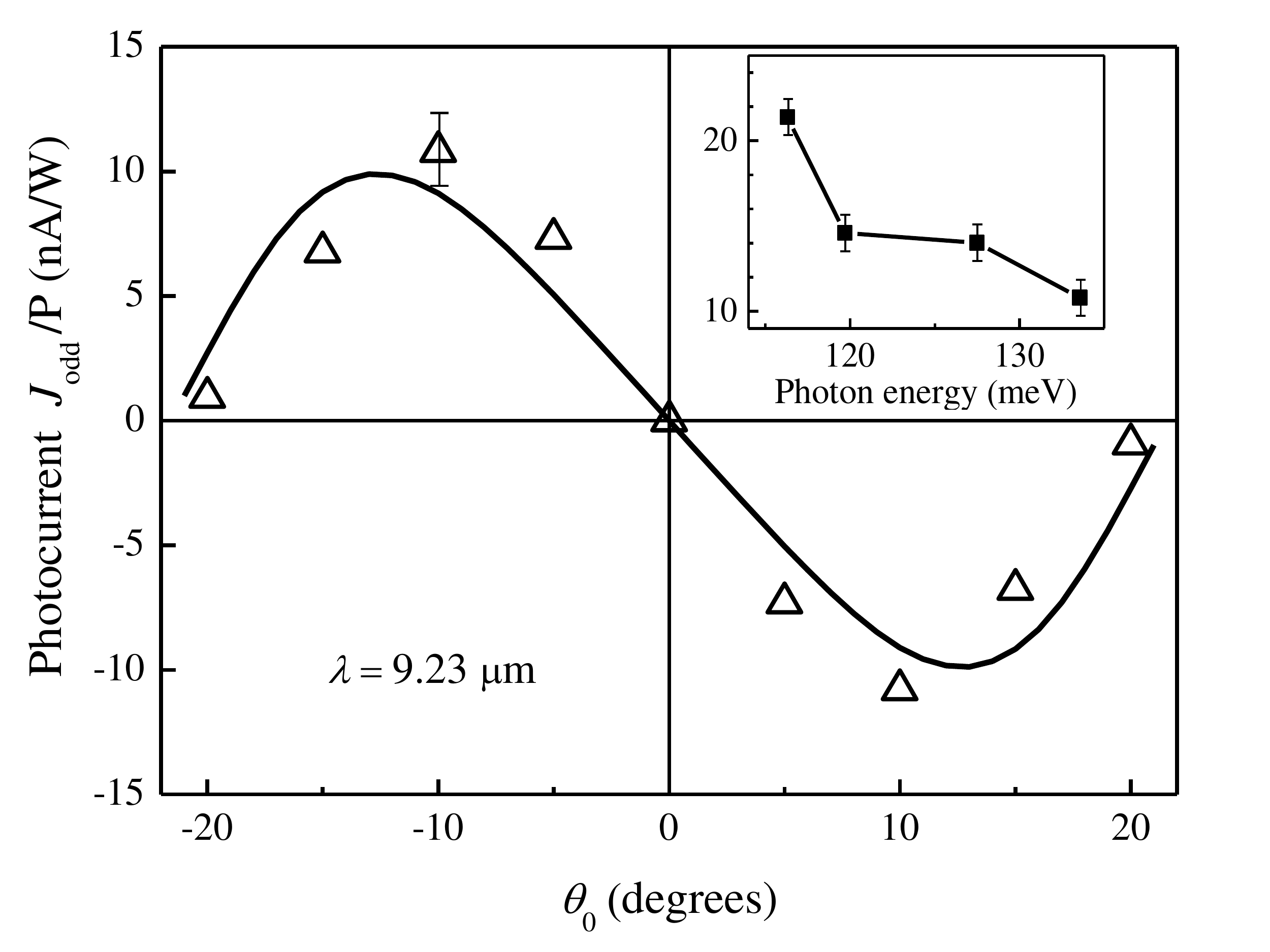}
\caption{Symbols: the circular photocurrent odd in the incidence angle~$\theta_0$.
The line is a fit by $\overline{j}(\theta_0)$ given by Eq.~\eqref{j1}. Inset shows the dependence of 
$|J_{\rm odd}|/P$ on the photon energy.
 }
\label{fig:odd}
\end{figure}

According to the phenomenological arguments, Sec.~\ref{Sec:phenom}, the CPDE current is an odd function of the incidence angle. Therefore we continued our analysis studying an  odd in $\theta_0$ part of $J_y^{circ}$. It is defined as follows:
\[
J_{\rm odd}(\theta_0) = {C(\theta_0)-C(-\theta_0)\over 2}.
\]
The dependence of  $J_{\rm odd}$ on the incidence angle  is plotted in Fig.~\ref{fig:odd}. 
We note that the helicity-dependent current exceeds by two orders of magnitude the current detected in quantum-well structures~\cite{CPDE_110_QWs,CPDE_110_QWs_2}.
We performed the same measurements and analysis at three other photon energies. 
The obtained dependence of $|J_{\rm odd}|$ on the photon energy is shown in inset in Fig.~\ref{fig:odd}. 

In order to verify that the odd in $\theta_0$ photocurent $J_{\rm odd}$ is caused by the CPDE, it is necessary to derive the dependence of CPDE on the incidence angle and to compare it with the experimental data. However it is not a trivial problem because tellurium is a birefringent crystal. It is characterized 
by two substantially different components of the dielectric susceptibility tensor 
$\varepsilon_\perp=23$ and $\varepsilon_\parallel=36$ for light polarized perpendicular and parallel to the optical axis $z$, respectively.
At an oblique incidence,
see Fig.~\ref{fig:exp_geom}, the light beam splits inside the medium into two, ordinary ($o$) and extraordinary ($e$) beams.
The $z$ components of the vectors $c\bm q/\omega$ for these beams are respectively given by
\begin{equation}
	n_o=\sqrt{\varepsilon_\perp-\sin^2{\theta_0}}, \qquad n_e =  \sqrt{\varepsilon_\perp-{\varepsilon_\perp\over \varepsilon_\parallel}\sin^2{\theta_0}}.
\end{equation}
Therefore the dependence of CPDE current on $\theta_0$ is a complicated function in contrast to the case of quantum-well structures, and to derive this dependence is an independent problem.

The transverse CPDE current Eq.~\eqref{j_phenom} can be equivalently presented as
\begin{equation}
\label{j_phenom_1}
	j_y^{circ} = T q_x {\rm i}(E_xE_y^* - E_yE_x^*).
\end{equation}
Since the ordinary and extraordinary beams propagate with different velocities, the CPDE current Eq.~\eqref{j_phenom_1} oscillates in space along the propagation direction like other helicity-dependent photocurrents in birefringent crystals~\cite{Belinicher78}. The period of these space oscillations is given by
\begin{equation}
	d(\theta_0) = {\lambda \over n_o -n_e},
\end{equation}
where $\lambda=2\pi c/\omega$ is the wavelength in vacuum.
The current density dependence on the coordinate $z$ inside the sample is the following:
\begin{equation}
	j_y^{circ}(z) = T q_x \cos{\left[{2\pi z \over d(\theta_0)}\right]} P_{circ}^0  {\cal T}_{ps}(\theta_0) E_0^2,
\end{equation}
where $E_0$ is the radiation amplitude in vacuum, and
the transmission coefficient is given by (see Appendix~\ref{AppendixA}) 
\begin{equation}
\label{T_ps}
	{\cal T}_{ps}(\theta_0) = 
	{4 n_e \cos^2{\theta_0} \over \left(\cos{\theta_0} + n_o\right) \left(\varepsilon_\perp\cos{\theta_0} + n_e \right)} .
\end{equation}
As a result, the CPDE current density in the sample of thickness $L$ depends on $\theta_0$ as follows:
\begin{multline}
\label{j1}
	{1\over L}\int\limits_0^L dz \, j_y^{circ}(z) \equiv \overline{j}(\theta_0) \\
	= T {\omega\over c} P_{circ}^0 E_0^2 {\cal T}_{ps}(\theta_0)\sin{\theta_0} \, { \sin{\left[2\pi L / d(\theta_0)\right]} \over 2\pi L / d(\theta_0)} .
\end{multline}

The absolute value of the experimentally detected photocurrent $|J_{\rm odd}(\theta_0)|$ is proportional to the average current density  given by Eq.~\eqref{j1}. Fit of the experimental data by the function $\overline{j}(\theta_0)$ is shown by a solid line in Fig.~\ref{fig:odd}.
One can see a good agreement between the theory and the experimental data.

The above analysis  of the photocurrent dependencies
on the laser power,  polarization state and incidence angle 
confirms observation of CPDE.

\section{Microscopic theory}
\label{Sec:theory}

Now we develop a microscopic theory of CPDE in tellurium. We derive the photon energy dependence of the CPDE current and compare it with the experimental data. 

\begin{figure}[h]
\includegraphics[width=0.9\linewidth]{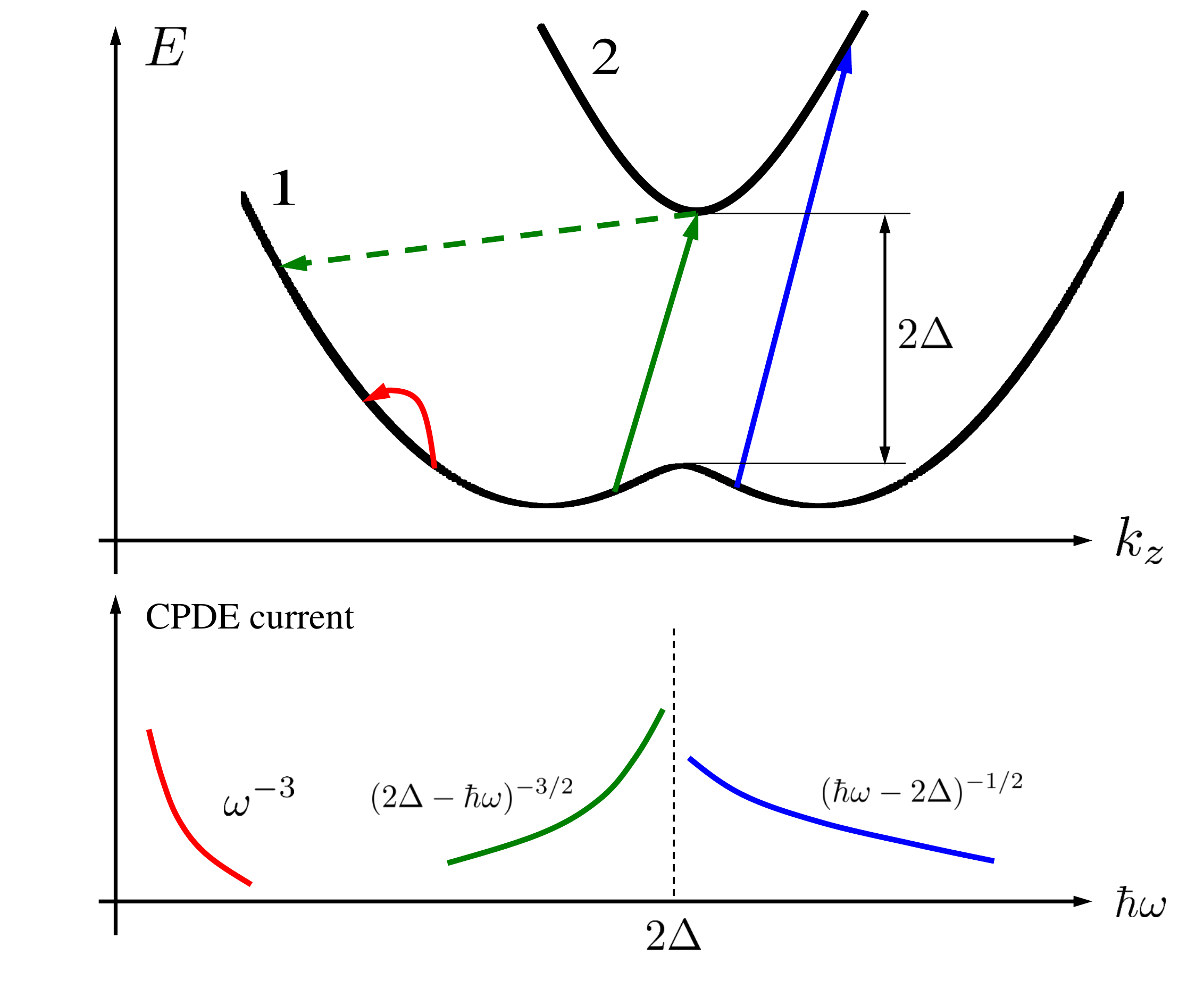}
\caption{Upper panel: The  valence band diagram of tellurium in the hole representation. A red arrow denotes optical intra-subband transition at low frequency, green arrows indicate a transition of the same type with an intermediate state in the excited subband accompanied by inter-subband scattering (dashed arrow), and a blue arrow shows the direct inter-subband optical transition. Lower panel: Frequency dependencies of the CPDE current absolute value in three frequency ranges corresponding to the transitions indicated in the upper panel.}
\label{fig:freq_depend_bandstruct}
\end{figure}

The effective Hamiltonian of holes in 
the tellurium valence band has the form~\cite{FTP1984}
\begin{equation}
\label{H}
	H({\bm k}) = Ak_z^2 + B k_\perp^2 + \beta \sigma_z k_z + \Delta \sigma_x,
\end{equation}
where $\bm k$ is a hole wavevector, and $\sigma_z$, $\sigma_x$ are the Pauli matrices.
The valence band is splitted at $k>0$ on two subbands with the dispersions
\begin{equation}
	E_{1,2}({\bm k}) = Ak_z^2 + B k_\perp^2 \mp\sqrt{\Delta^2 + \beta^2 k_z^2},
\end{equation}
which are plotted in Fig.~\ref{fig:freq_depend_bandstruct}.

We calculate the CPDE constant $T$ in Eqs.~\eqref{j_phenom},~\eqref{j_phenom_1} for both intra-subband and inter-subband absorption in the tellurium valence band.
 
\subsection{Drude-like absorption}

At low frequencies $\hbar\omega \ll  \Theta$, where $\Theta$ is the temperature in energy units, light absorption is caused by intraband transitions. The CPDE current at Drude-like  absorption is found by solving the Boltzmann kinetic equation.
In order to get
the CPDE photocurrent, we take into account both coordinate dependence of the distribution function and the
Lorentz force caused by the magnetic field $\bm B$ of the light wave. In the relaxation-time approximation,
the Boltzmann equation for the distribution function $f$ dependent on the hole wavevector $\bm k$, coordinate $\bm r$ and time $t$ has the form
\begin{equation}
	{\partial f \over \partial t} + \bm v_{\bm k} \cdot {\partial f \over \partial \bm r} + 
	{e\over \hbar}\bm E \cdot {\partial f \over \partial \bm k} + {e\over \hbar c} (\bm v_{\bm k} \times \bm B) \cdot {\partial f \over \partial \bm k} = - {f - \bar{f} \over \tau}.
\end{equation}
Here the bar denotes averaging over directions of $\bm k$, $\tau$ is the momentum relaxation time, and $\bm v_{\bm k}=\hbar^{-1}\partial \varepsilon_{\bm k}/\partial \bm k$ is the hole velocity with $\varepsilon_{\bm k} \equiv E_1(\bm k)$ being the hole energy dispersion in the ground valence subband.
The photocurrent density is given by
\begin{equation}
	\bm j = 2e\sum_{\bm k} \bm v_{\bm k} \left<f(\bm k,\bm r,t)\right>,
\end{equation}
where the factor 2 accounts for two tellurium valleys, and angular brackets mean averaging over both space coordinate and time.

Solving the Boltzmann equation by iterations in the second order in $\bm E$ and in the first order in the space gradient at $B=0$, we find the first part of the CPDE current. Taking into account both $\bm E$ and $\bm B$ in the first orders but ignoring the coordinate dependence, we get the second part~\cite{graphene_ac_Hall,GlazovGanichevReview}. 
As a result, we obtain the CPDE current density in the form:
\begin{multline}
\label{Drude}
j_y^{circ} = q_x {\rm i}(E_xE_y^* - E_yE_x^*) \\
	\times 2e^3 \sum_{\bm k} {d\tau \over d \varepsilon_{\bm k}} {\tau^2 (-df_0/d\varepsilon_{\bm k}) \over [1+(\omega\tau)^2]^2} v_\perp^2 \left({v_\perp^2\over 2} - {\varepsilon_\perp \over \varepsilon_\parallel} v_z^2\right) .
\end{multline}
 Here $f_0(\varepsilon_{\bm k})$ is the Boltzmann distribution function, and $v_\perp^2=v_x^2+v_y^2$. Deriving this expression we have taken into account the relation ${ div} \bm D =0$ which yields
 \begin{equation}
\label{epsilons}
\varepsilon_\parallel q_zE_z + \varepsilon_\perp \bm q_\perp \cdot \bm E_\perp=0.	 
 \end{equation}

\subsection{Inter-subband transitions}

At photon energy larger than the inter-subband gap, $\hbar\omega > 2\Delta$, a light absorption is caused by direct optical transitions. The CPDE current can be calculated in this case by quantum-mechanical methods. The analysis shows that the CPDE current is a sum of two terms. The first, \textit{ballistic} contribution arises from account for an additional scattering process side by side with the optical transition~\cite{Belinicher_1981}. The second contribution called \textit{shift} photocurrent is caused by shifts of carriers in the real space occurring at photon absorption~\cite{EL_book}. We estimate the value of CPDE current calculating the shift contribution.

The shift photocurrent density at direct optical transitions is given by
\begin{multline}
\label{j_direct}
	{\bm j}^{inter} = -2e\sum_{{\bm k}} {2\pi \over \hbar} \delta[E_2({\bm k} + {\bm q}) -E_1({\bm k}) - \hbar\omega] \\
	 \times \{f_0[E_1({\bm k})] - f_0[E_2({\bm k} + {\bm q})]\} \, {\rm Im} (V_{21}^* \nabla_{\bm k} V_{21}), 
\end{multline}
where 
$V_{21}$ is a matrix element of the operator of direct optical transitions accounting for the photon wavevector $\bm q$.
The wavefunction envelopes are eigenvectors of the Hamiltonian~\eqref{H}:
${\Psi_1 = (C_1,C_2)}$, ${\Psi_2 = (C_2,-C_1)}$,
where ${C_{1,2} = \sqrt{(1 \pm \eta)/2} }$, ${\eta=\beta k_z/\sqrt{\Delta^2 + \beta^2 k_z^2} }$~\cite{FTP1984}.
The direct optical transitions at $q=0$ are allowed in the polarization $\bm E \parallel z$ only. However, with account for the photon momentum, the polarization $\bm E_\perp \perp z$ also interacts with the carriers:
\begin{equation}
\label{V_21}
	V_{21} = {{\rm i}e \beta  \Delta \over \hbar \omega}  
	 \left( {E_z \over \sqrt{\Delta^2 + \beta^2 k_z^2} } - B q_z {\bm E_\perp \cdot \bm k_\perp \over \Delta^2 + \beta^2 k_z^2} \right).
\end{equation}
It follows from this equation that the coordinate shifts of holes in the  $y$ direction 
proportional to the light helicity are present at $q \neq 0$:
\begin{equation}
	\Delta y \sim -{{\rm Im} (V_{21}^* \partial V_{21}/\partial k_y) \over |V_{21}|^2} \sim P_{circ} {B q \over \sqrt{\Delta^2 + \beta^2 k_z^2} }.
\end{equation}

Calculation by Eq.~\eqref{j_direct} with account for the relation~\eqref{epsilons}
leads to the CPDE current in the form 
\begin{align}
\label{T_direct}
& j_y^{circ} = - q_x {\rm i}(E_xE_y^* - E_yE_x^*) 
	  {8\pi p e^3 B |\beta| \Delta^2 \sqrt{A} \varepsilon_\perp\over J \hbar^5 \omega^4 \sqrt{\Theta} \varepsilon_\parallel}
	   \\
& \times	\left(1 - {\rm e}^{-\hbar\omega/\Theta} \right) { {\rm exp}\left\{ - {A \over 4\beta^2\Theta} [(\hbar\omega)^2 - (2\Delta)^2] + {\hbar\omega-2\Delta\over 2\Theta}\right\}
	\over \sqrt{(\hbar\omega)^2 - (2\Delta)^2} }.	\nonumber
\end{align}
Here $p$ is the hole concentration, and
\begin{equation}
\label{J}
	J =  \int\limits_0^\infty dy \exp{\left(-y^2 - {\Delta \over \Theta} \right)}
	 \, \cosh\left( {\sqrt{\Delta^2 + {\beta^2\Theta\over A}y^2} \over \Theta}\right).
\end{equation}

\subsection{Intra-subband transitions via the excited subband}

At photon energy smaller than but comparable to the gap between the valence subbands, $\hbar\omega \leq 2\Delta$, optical absorption is caused by two-step transitions with an intermediate state in the excited subband, see Fig.~\ref{fig:freq_depend_bandstruct}. The shift current density in this case is given by~\cite{JETP_shift}
\begin{align}
&	{\bm j}^{intra} = -2e\sum_{{\bm k}, {\bm k'}} {2\pi \over \hbar} \delta[E_1({\bm k}) -E_1({\bm k'}) - \hbar\omega] \\
& \times	  \{f_0[E_1({\bm k})] - f_0[E_1({\bm k'})]\} \, {{\rm Im} [W^* (\bm \nabla_{\bm k}+ \bm \nabla_{\bm k'}) W] \over
		[E_1(\bm k) + \hbar\omega - E_2(\bm k+ \bm q)]^2}, \nonumber
\end{align}
with  $W$ being a product:
\begin{equation}
W = U^{12}_{\bm k', \bm k+ \bm q} V_{21}.
\end{equation}
Here $V_{21}$ is a matrix element of direct optical transitions~\eqref{V_21} introduced in the previous subsection, and $U^{12}_{\bm k', \bm k+ \bm q}$ is the 
 matrix element of  intersubband scattering from the state $(2,\bm k+ \bm q)$ to the state $(1,\bm k')$ shown by the dashed line in Fig.~\ref{fig:freq_depend_bandstruct}. We assume the scattering to be elastic. Here we account only  for the resonant term in the two-step transition matrix element which exceeds the other, non-resonant one, at $\hbar\omega \approx 2\Delta$.

Since the wavefunction envelopes are independent of $\bm k_\perp$, the scattering amplitude $U^{12}_{\bm k', \bm k+ \bm q}$ can depend on the difference $\bm k'_\perp - \bm k_\perp$ only. Hence, calculating the $y$ photocurrent component, we should differentiate in $W$
only the optical matrix element $V_{21}$. It is given by Eq.~\eqref{V_21}, therefore the elementary shifts coincide with those  at  direct intersubband transitions.
Introducing the scattering time $\tilde{\tau}$  according to
\begin{equation}
	{1 \over \tilde{\tau}}=	\sum_{\bm k'} {2\pi \over \hbar} \left|U^{12}_{\bm k', \bm k} \right|^2 \delta[E_1({\bm k}) -E_1({\bm k'}) - \hbar\omega] 
\end{equation}
and neglecting a dependence of $\tilde{\tau}$ on $\bm k$ and $\omega$,
we obtain the photocurrent in the form:
\begin{align}
\label{T_indirect}
& j_y^{circ} = - q_x {\rm i}(E_xE_y^* - E_yE_x^*) 
	{p \, e^3 B |\beta| \sqrt{A} \varepsilon_\perp \over 4J\Delta^2(\hbar \omega)^2 \tilde{\tau}\sqrt{\Theta}\varepsilon_\parallel}
	 \\
& \times	\left(1 - {\rm e}^{-\hbar\omega/\Theta} \right) \int\limits_0^\infty dx	
	{\exp{\left[ - {A\Delta^2\over\beta^2\Theta}x^2 + {\Delta\over \Theta}(\sqrt{1+x^2}-1) \right]}
	\over (\sqrt{1+x^2} - \hbar\omega/2\Delta)^2 (1+x^2)^{3/2}}. \nonumber
\end{align}
Here $J$ is introduced in Eq.~\eqref{J}.

The calculated CPDE current  dependence on the light polarization at Drude-like, inter- and intra-subband optical transitions, Eqs.~\eqref{Drude},~\eqref{T_direct} and~\eqref{T_indirect}, is in agreement with the phenomenological theory, Eqs.~\eqref{j_phenom},~\eqref{j_phenom_1}.

\section{Discussion}
\label{Sec:discussion}

The developed theory of CPDE allows us to describe all experimental findings. In particular, both phenomenological and microscopic theory yield the experimentally observed dependence on the incidence angle, see Eq.~\eqref{j1} and the fit in Fig.~\ref{fig:odd}. 
However, despite Fig.~\ref{fig:phi-dependence}(b) demonstrates that the circular photocurrent is mainly an odd function of $\theta_0$, an admixture of a small even contribution is present. In ideal tellurium, an even in $\theta_0$ transverse photocurrent is forbidden by symmetry, and its presence in the studied sample is caused by some asymmetry in the $xy$-plane. The cross-section of the grown single crystal indeed does not represent a regular hexagon. The adjacent sides of hexagonal cross-section had different lengths, namely 5~mm and 3~mm (see Fig.~\ref{fig:exp_geom}). This asymmetry appears in the process of growth: at stretching of samples, an 
inhomogeneous distribution of the diameter in the $z$-direction is introduced due to  pulsations of a heater. 
The non-ideal samples have the point symmetry group  C$_1$. In this case, the even in $\theta_0$ helicity-dependent currents are allowed due to both CPDE and the circular photogalvanic effect.
A polarization-independent contribution to the transverse photocurrent [the term $L_2$ in Eq.~\eqref{E1}] 
is also caused by the above mentioned asymmetry.
Nevertheless the dominating odd in $\theta_0$ contribution is well described by the developed theory which confirms observation of the CPDE.

Due to birefringence of tellurium, see Sec.~\ref{Sec:observation}, the light circularly polarized in vacuum became elliptically polarized inside the sample. A presence of a linear polarization can lead to additional photogalvanic currents caused by both the linear photon drag and the linear photogalvanic effects. The corresponding odd in $\theta_0$ contribution behaves as $\propto P_{circ}^0\theta_0^3$ at small incidence angles. Our estimate shows that it does not exceed 20~\% of the CPDE contribution Eq.~\eqref{j_phenom}. 

The CPDE current is an even function of the constant $\beta$, see Sec.~\ref{Sec:theory}. 
This result contrasts to the photocurrent caused by the circular photogalvanic effect which is linear in $\beta$. The reason is that CPDE is insensitive to a presence of a space inversion center and therefore is the same in laevorotatory and dextrorotatory  tellurium which are different by a sign of constant $\beta$.

Microscopic theory also describes
the measured photon energy dependence presented in inset to Fig.~\ref{fig:odd}. 
The frequency dependencies of the transverse CPDE photocurrent for all three types of transitions considered above are shown in Fig.~\ref{fig:freq_depend_bandstruct}.
At $\hbar \omega \ll \Theta < 2\Delta$, the frequency dependence is given by Eq.~\eqref{Drude}. This equation demonstrates that CPDE is present due to energy dependence of the momentum relaxation time. Besides, it is also necessary to account for a uniaxiality of tellurium, because CPDE is forbidden in systems of cubic symmetry. It follows from Eq.~\eqref{Drude} that CPDE in tellurium exists due to anisotropy of the energy spectrum and the dielectric tensor.  At $\omega\tau \ll 1$, 
the CPDE current increases linearly  with frequency, $j_y^{circ}\propto \omega$. At $\omega\tau \gg 1$ but still $\hbar \omega \ll \Theta$, the current decreases as $j_y^{circ}\propto \omega^{-3}$. Both these asymptotes coincide with the high-frequency behavior of CPDE current in graphene~\cite{GlazovGanichevReview}.

Near the direct absorption edge, $\hbar \omega = 2\Delta$, the CPDE current has a singularity corresponding to the transitions from the states with $k_z=0$. It follows from Eq.~\eqref{T_direct} that above the direct absorption edge, $\hbar \omega \to 2\Delta +0$, the coefficient $T$ in the phenomenological Eq.~\eqref{j_phenom_1} has a square-root singularity:
\begin{equation}
\label{T_direct_sing}
	T_{inter}
	= - {p e^3 B |\beta| \sqrt{A} \pi \varepsilon_\perp\over 4J\hbar \Delta^{5/2}\sqrt{\Theta}\varepsilon_\parallel}
	{1 - {\rm e}^{-2\Delta/\Theta} \over \sqrt{\hbar\omega - 2\Delta}}.
\end{equation}
This singularity is one dimensional optical density of states which arises at direct optical transitions between the valence subbands. This van Hove singularity is not integrated because of equal energy dependencies on the perpendicular wavevector $k_\perp$ in both valence subbands. 

Below the direct absorption edge,  
the singularity at $\hbar \omega \to 2\Delta -0$ follows from the  contribution to the integral Eq.~\eqref{T_indirect} from $x=0$ which yields
\begin{equation}
\label{T_indirect_sing}
	T_{intra}
	= - {\pi p \, e^3 B |\beta| \sqrt{A} \varepsilon_\perp \over 16 J \Delta^{5/2} \tilde{\tau} \sqrt{\Theta}\varepsilon_\parallel}
	 {1 - {\rm e}^{-2\Delta/\Theta} \over (2\Delta-\hbar\omega)^{3/2}}
	.	
\end{equation}
We see that the singularity is stronger  at intra-band transitions than at inter-band transitions by a factor ${\sim {\hbar / (|2\Delta-\hbar\omega|\tilde{\tau})} \gg 1}$.

Experimental data demonstrates a decrease of the CPDE current $|J_{\rm odd}|$ with photon energy, see inset in Fig.~\ref{fig:odd}. Theoretically, we obtain a decrease of the current with frequency at direct inter-subband transitions (for $\hbar\omega>2\Delta$) illustrated in Fig.~\ref{fig:freq_depend_bandstruct}. This explains qualitatively the spectral behavior of CPDE current observed in experiment.
We have estimated the CPDE current at inter-subband transitions with help of Eqs.~\eqref{j1},~\eqref{T_direct_sing}. For tellurium at room temperature $A=3.71 \times 10^{-15}$~eVcm$^2$, $B=3.57 \times 10^{-15}$~eVcm$^2$, $\beta=2.5 \times 10^{-8}$~eVcm, $2\Delta=126$~meV~\cite{FTP1984_II},
we obtain {$\overline{j}/P \approx 50$~nA/W} at $\hbar\omega=130$~meV and $\theta_0=10^\circ$. This is the same order of magnitude as the experimental data, see Fig.~\ref{fig:odd}. A bit smaller values of the current ($\sim 10$~nA/W) are detected in the experiment because not all current generated in the laser spot area reaches the contacts. A part of the current is closed in the non-illuminated part of the sample.

\section{Conclusion}
\label{Sec:conclusion}

In conclusion, the helicity-dependent photocurrent transverse to the light incidence plane is detected in bulk tellurium. 
The above analysis of the polarization state, incidence angle and photon energy dependencies of the photocurrent confirms observation of the CPDE.
The  CPDE current is shown to be an odd function of the incidence angle.
The phenomenological model of CPDE is developed based on symmetry arguments with account for birefringence of tellurium. 
Microscopic theory for both inter- and intra-subband optical transitions is elaborated. The CPDE current is estimated by calculation of the shift contribution. The resonance  in CPDE current frequency dependence at the threshold of the intersubband transitions is demonstrated theoretically.
Theory yields the same photon energy dependence and the values of the circular photocurrent as in the experiment.
Due to high sensitivity of CPDE, tellurium  
can be used for helicity-dependent photodetectors. This opens a way for all-electric detection of a light polarization state.
Finally we note that CPDE also can be present in topological insulators based on tellurides, and its study can be helpful for understanding of their symmetry and kinetic properties.

\appendix
\section{Transmision coefficient to a uniaxial medium} 
\label{AppendixA}

The amplitude of the ordinary beam $\bm E \parallel y$ at the boundary with vacuum ($z=0$) is given by the Fresnel transmission coefficient for $s$-polarized light: 
\begin{equation}
	E_y(z=0) = {2\cos{\theta_0} \over \cos{\theta_0} + n_o } E_{0y}.
\end{equation}
Here $\bm E_0$ is the radiation amplitude in vacuum.

Under incidence of $p$-polarized light, the extraordinary beam is excited. From the Maxwell boundary conditions we obtain the amplitude of the transmitted wave at ${z=0}$. For its $x$ component we have:
\begin{equation}
	E_x(z=0) = {2 n_e \over n_e + \varepsilon_\perp \cos{\theta_0} } E_{0x}.
\end{equation}
As a result, we have for bilinear combinations:
\begin{equation}
	E_x E_y = {4 n_e \cos{\theta_0}  \over \left(\cos{\theta_0} + n_o\right)
	\left(\varepsilon_\perp\cos{\theta_0} + n_e \right) } E_{0x} E_{0y}.
\end{equation}

Since the circular polarization degree in vacuum $P_{circ}^0$ is defined via
\begin{equation}
	{\rm i}(E_{0x} E_{0y}^*-E_{0y}^0E_{0x}^*) = P_{circ}^0 E_0^2 \cos{\theta_0},
\end{equation}
the same combination for the transmitted light at the boundary with vacuum has the form
\begin{equation}
	{\rm i}(E_xE_y^{*}-E_yE_x^{*})|_{z=0} = {\cal T}_{ps} P_{circ}^0 E_0^2,
\end{equation}
with ${\cal T}_{ps}$ given by Eq.~\eqref{T_ps}.

\section*{Acknowledgments}
We thank S.~D.~Ganichev and E.~L.~Ivchenko for helpful discussions, I.~V.~Sedova for assistance in sample preparation, N.~V.~Ageev for help in transport measurements, and V.~K. Kalevich for interest to the work.
Financial  support from the Russian Foundation for
Basic Research, Ministry of Education and Science of the Russian Federation and  DFG (SPP 1666) is gratefully acknowledged. The work of L.~E.~G. was supported by the Russian Science Foundation (project 14-12-01067).

\end{document}